\documentclass[sn-basic,Numbered]{sn-jnl}%

\usepackage{graphicx}%
\usepackage{multirow}%
\usepackage{amsmath,amssymb,amsfonts}%
\usepackage{amsthm}%
\usepackage{mathrsfs}%
\usepackage[title]{appendix}%
\usepackage{xcolor}%
\usepackage{textcomp}%
\usepackage{manyfoot}%
\usepackage{booktabs}%
\usepackage{algorithm}%
\usepackage{algorithmicx}%
\usepackage{algpseudocode}%
\usepackage{listings}%
\usepackage[framemethod=TikZ]{mdframed}
\usepackage{orcidlink}

\usepackage{tcolorbox}
\tcbuselibrary{theorems}
\usepackage{fontawesome}

\newtcbtheorem{takeaway}{\faLightbulbO{} Takeaway}
{colback=green!5,colframe=green!35!black,fonttitle=\bfseries}{th}

\newcounter{lem}[section]
\renewcommand{\thelem}{\arabic{section}.\arabic{lem}}

\newcommand{\sectopic}[1]{\vspace{0.2em}\par\noindent{\textit{\bfseries #1}}}

\begin{document}

\title[15. Practical Guidelines for the Selection and Evaluation of Natural Language Processing Techniques in Requirements Engineering]{15. Practical Guidelines for the Selection and Evaluation of Natural Language Processing Techniques in Requirements Engineering}

\author[1]{\fnm{Mehrdad} \sur{Sabetzadeh}\orcidlink{0000-0002-4711-8319}}\email{m.sabetzadeh@uottawa.ca}

\author[2]{\fnm{Chetan} \sur{Arora}\orcidlink{0000-0003-1466-7386}}\email{chetan.arora@monash.edu}

\affil[1]{
\orgname{University of Ottawa}, 
\state{ON}, \country{Canada}}

\affil[2]{
\orgname{Monash University}, 
\state{Vic}, \country{Australia}}

\abstract{Natural Language Processing (NLP) is now a cornerstone of requirements automation. One compelling factor behind the growing adoption of NLP in Requirements Engineering (RE) is the prevalent use of natural language (NL) for specifying requirements in industry. NLP techniques are commonly used for automatically classifying requirements, extracting important information, e.g., domain models and glossary terms, and performing quality assurance tasks, such as ambiguity handling and completeness checking. With so many different NLP solution strategies available and the possibility of applying machine learning alongside, it can be challenging to choose the right strategy for a specific RE task and to evaluate the resulting solution in an empirically rigorous manner. In this chapter, we present guidelines for the selection of NLP techniques as well as for their evaluation in the context of RE. In particular, we discuss how to choose among different strategies such as traditional NLP, feature-based machine learning, and language-model-based methods. Our ultimate hope for this chapter is to serve as a stepping stone, assisting newcomers to NLP4RE in quickly initiating themselves into the NLP technologies most pertinent to the RE field.}

\maketitle

\section{Introduction}
NLP's role in requirements automation is pivotal, due to the widespread use of natural language (NL) in industrial requirements specifications. Historically, NL has posed challenges for requirements analysis because of its inherent proneness to defects such as incompleteness and ambiguity. Recent breakthroughs in NLP, e.g., the emergence of large language models, have nonetheless drastically enhanced our ability to automatically analyze textual information. This development is poised to even further amplify the adoption and influence of NL in requirements engineering.

Due to the rapid advancement of NLP, newcomers to NLP4RE may feel overwhelmed by the numerous potentially applicable technologies. Another challenge is the necessity to empirically assess a proposed automation solution, ensuring proper optimization, and, where applicable, improve performance over existing solutions. 

Over the past several years, we have studied various requirements automation problems, including checking compliance with requirements templates~\cite{Arora:15a}, glossary construction~\cite{Arora:17}, model extraction~\cite{Arora:16}, requirements demarcation~\cite{Abualhaija:20b}, ambiguity handling~\cite{Ezzini:22}, and question answering~\cite{Ezzini:23}. With the benefit of hindsight, this chapter aims to reflect on our research process and offer our collective insights into how we approach NLP4RE problems. Before we start, we need to emphasize that our perspective is \emph{retrospective}. Given the fast pace of progress in NLP technologies, new considerations may surface, and existing technologies could become outdated. Therefore, it is important for readers to consider the time this chapter was written (2023) when dealing with new technologies. This advice applies to most works in the fast-changing field of Applied AI.

\sectopic{Structure.}
Section~\ref{sec:AutomationSteps} outlines the steps for automating pre-processing, analysis, and post-processing in NLP4RE. Section~\ref{sec:enablingTechniques} describes various NLP techniques and discusses their key considerations for automation in RE. Finally, Section~\ref{sec:conclusion} summarizes the chapter and presents conclusions.
\section{Automation Steps in NLP4RE}~\label{sec:AutomationSteps}
\begin{figure}
    \centering
    \includegraphics[width=\textwidth]{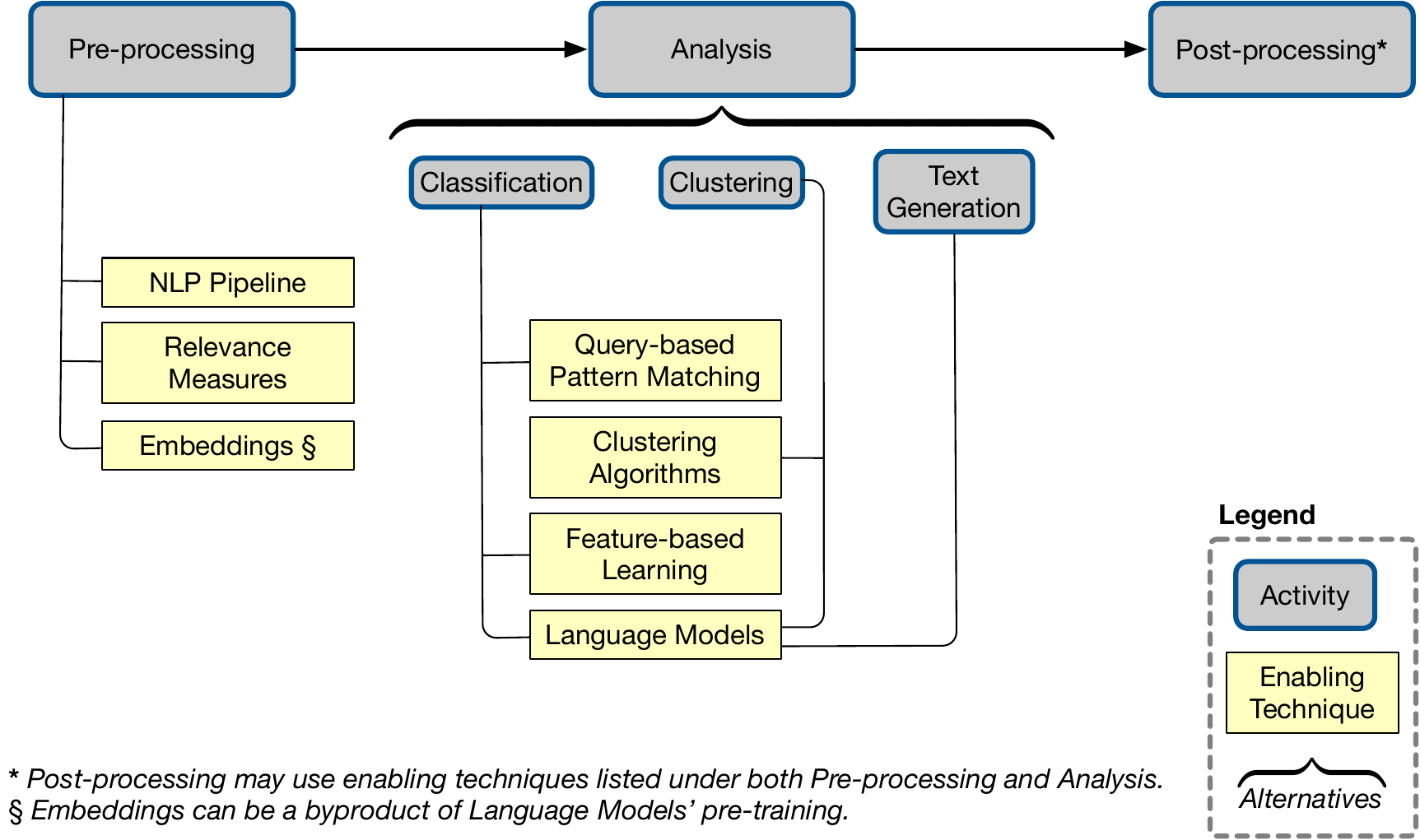}
    \caption{NLP4RE Steps}
    \label{fig:overview}
\end{figure}

The automation process for NL requirements and related textual artifacts can be structured into three sequential steps. These steps, shown in Figure~\ref{fig:overview} are: (1) Pre-processing, (2) Analysis, and (3) Post-processing. We outline these steps next.

\subsection{Pre-processing}\label{subsec:pre}
The goal of pre-processing is to automatically examine the NL content of requirements or requirements-related artifacts (e.g., design documents and code) and generate structured information for use by the Analysis step (Section ~\ref{subsec:analysis}).

The specific information targeted by pre-processing depends on the needs of the subsequent Analysis step. Importantly, the information to obtain through pre-processing relies on the selection of the \emph{units of analysis} for the Analysis step. In the context of NLP, ``unit of analysis'' refers to the particular textual components that are intended for processing and interpretation. Some common units of analysis are: words, phrases, sentences, and paragraphs. For instance, when the objective of Analysis is identifying statements that contain requirements, sentences are frequently adopted as the units of analysis~\cite{Abualhaija:20b}. We note that there is a composition relationship between different unit types in NLP. For example, a sentence is made up of phrases, and a phrase is made up of words. Because of this characteristic, it is possible to use a combination of units of analysis simultaneously, e.g., phrases alongside sentences. 

Typically, during the pre-processing phase, a set of ``features'' are calculated for the units of analysis (or the relationships between the units). A \emph{feature} refers to a distinct attribute or characteristic of a unit of analysis. Features can be numeric (e.g., the number of tokens appearing in a sentence) or categorical (e.g., part-of-speech tags).

The most common enabling technologies for computing features are presented in Figure~\ref{fig:overview} under the Pre-processing activity. These technologies include the NLP Pipeline, Relevance Measures and  Embeddings, which are discussed further in Section~\ref{sec:enablingTechniques}.

\subsection{Analysis}\label{subsec:analysis}
The core of the process shown in Figure~\ref{fig:overview} is the Analysis step. In NLP4RE, this step typically manifests as one of the following three alternative activities: Classification, Clustering, and Text Generation.

\subsubsection{Analysis Activities}
\sectopic{Classification}
involves the assignment of labels or categories to the units of analysis. There are numerous NLP4RE use cases for classification. An example use case would be differentiating between functional (F) and non-functional requirements (NF)~\cite{Kurtanovic:17}. This task can be framed as the assignment of F and NF labels to the units of analysis, which in this context, are typically sentences within a requirements document.

Classification further extends to encompass \emph{verbatim information extraction}. Verbatim information extraction involves directly extracting exact segments from source documents without abstraction, inference, or interpretation. This is done by marking off the text segments of interest and assigning labels to them. A typical use case is identifying requirements-related text segments in legal documents and annotating them with labels such as ``permission'', ``obligation'', ``condition'', and ``exception''~\cite{Sleimi:21}.
\vspace*{0.5em}

\sectopic{Clustering} leverages inherent similarities among the units of analysis to organize them into groups or themes. Unlike classification, which requires predetermined labels, clustering  emphasizes the intrinsic structure and similarities within the content. The goal is to bring together similar content based on shared features, thereby avoiding the need for explicit predefined categories. This makes clustering particularly useful when dealing with problems where the labels are not well-defined or when exploring content where the underlying patterns might not be immediately evident. For instance, clustering can be used to identify groups of closely related requirements phrases during glossary construction~\cite{Arora:17}; here there is no predetermined set of classes for the groups of related phrases that will emerge.

\vspace*{0.5em}
\sectopic{Text Generation} involves the automated creation of human-readable text based on either structured or unstructured inputs to aid the derivation, completion, understanding and communication of requirements. NLP4RE solutions based on text generation are a relatively recent development but are rapidly gaining momentum thanks to advances in generative language models like GPT~\cite{Arora:23}. Example use cases for text generation include constructing requirements models based on prompts and early-stage descriptions~\cite{Chen:23}, summarizing requirements-related documents~\cite{Jain:23}, and providing predictive assistance for requirements completion~\cite{Luitel:23}.

\subsubsection{Technique Selection}
\begin{figure}
    \centering
\includegraphics[width=\textwidth]{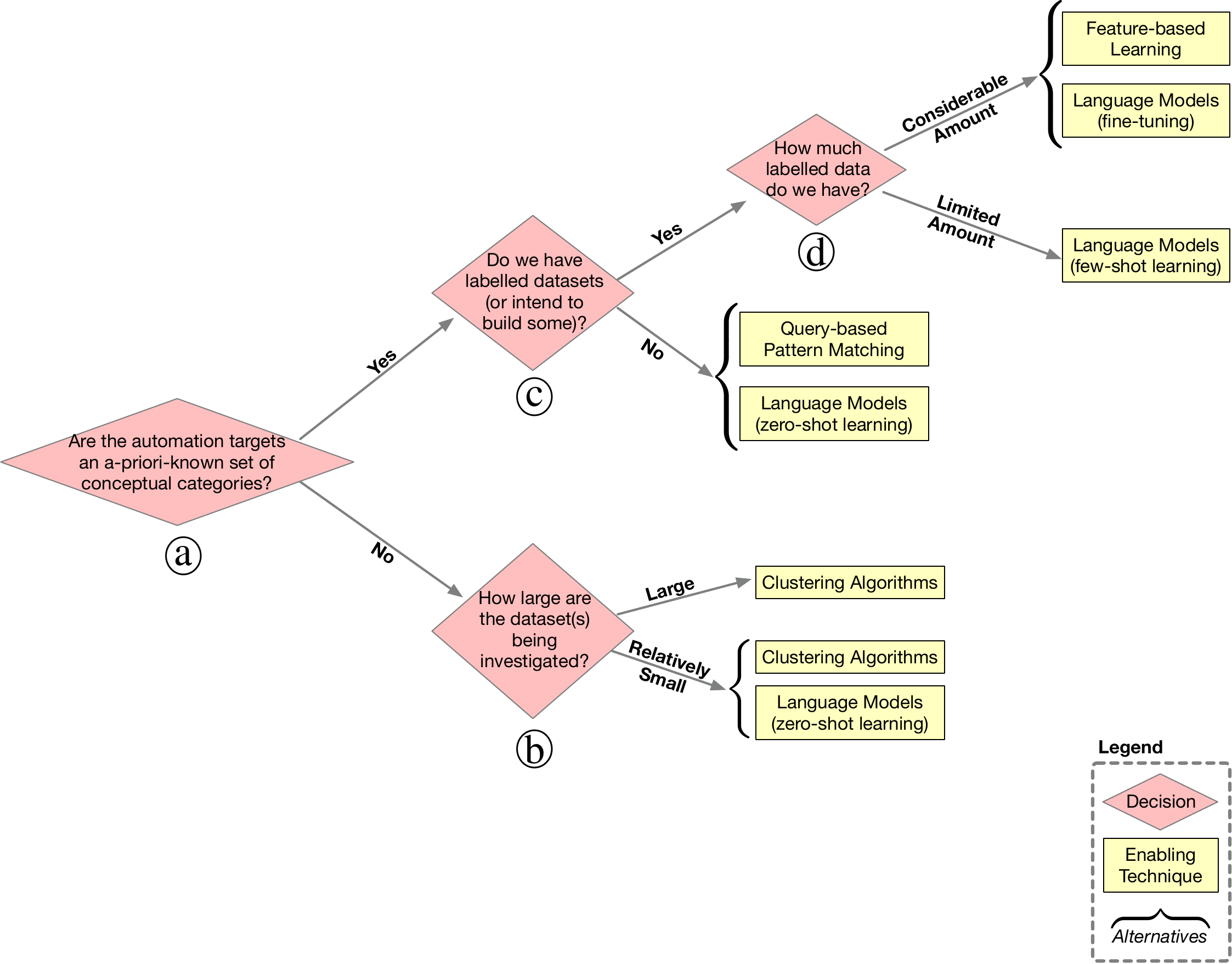}
    \caption{Identifying Suitable Enabling Technique(s) for a Specific Analysis Task}
    \label{fig:decision}
\end{figure}

Selecting suitable enabling technique(s) for the Analysis step is crucial. To make this selection easier, we have developed a decision process, shown in Figure~\ref{fig:decision}. This simple process, which is based on our past experience, aims to facilitate narrowing the options for the Analysis step.

The first and most important criterion in this process is decision node~(a), as depicted in Figure~\ref{fig:decision}. This decision concerns whether we have a well-established and pre-existing set of conceptual categories relevant to automation. For instance, consider the task of classifying functional and non-functional requirements. In this task, the categories would be functional and non-functional, and this understanding exists before classification. For another example, consider the task of domain model extraction. Here, all pertinent categories are identified and can be listed, such as class, attribute, association, cardinality constraint, and so on. In contrast, consider a problem like identifying similar and potentially redundant requirements. When presented with a requirements document, predicting the number of equivalence classes (clusters) is impossible. As a result, there is no predefined set of conceptual categories for this problem that can be known beforehand.

Without established conceptual categories, dataset size is the next crucial criterion to consider. Historically, when predefined categories are not possible, clustering algorithms have been the preferred enabling technique for analysis. Recent advances in generative language models like ChatGPT and Llama nonetheless offer an alternative. For instance, when dealing with data sources such as requirements documents, one can use exploratory prompts like the following to uncover the main themes: \emph{``Cluster the primary concepts in the following: [contents of the data source]''}.

As of writing, using a generative language model as an alternative to clustering algorithms is effective only for relatively small content volumes. Existing language models set a limit on total input and output tokens during interactive dialogs, known as ``token limit'' or ``session size.'' This limit defines how much context the model is capable of taking into account for response generation. Currently, the largest token limit we are aware of is 32,768 tokens (GPT-4-32k); this would be insufficient for many NLP4RE problems, such as traceability retrieval, which can potentially involve millions of tokens.

When an a-priori-known set of conceptual categories exists, the next question is whether a labelled dataset is available. This is captured by decision node (c) in Figure~\ref{fig:decision}. 
When a labelled dataset is lacking, the most common enabling technique is query-based pattern matching. Although pattern matching does not require the explicit creation of labelled data, formulating the queries often entails some form of qualitative analysis, e.g., grounded theory~\cite{Saldana:15}. Alternatively, language models can be employed when labelled data is unavailable. In such a scenario, one has to rely exclusively on the language model's pre-training, without further specific training for the task at hand. For example, a language model could be asked the following: ``Is this requirements statement functional or non-functional? \emph{[an individual requirements statement]}''.

If labelled data is available, the next factor to consider is the volume of such data, as captured by decision node~(d) in Figure~\ref{fig:decision}. 
When a considerable amount of labelled data is available, there are two options: applying feature-based learning or utilizing the labelled data to ``fine-tune'' a language model. Fine-tuning involves adapting a pre-trained model to a specific task through targeted training. In cases of limited labelled data, achieving task adaptation for a language model with minimal examples -- a technique known as few-shot learning -- is likely to yield better results.

It is important to note that there is no general rule as to what constitutes ``considerable'' or ``limited'' labelled data. Several parameters such as the quality of labels, the complexity of the relationships to be learned, the number of classes (in classification) and the range of values (in regression), the amount of noise in the data, the dimensionality of the feature space (in case of feature-based learning), and the desired level of accuracy to achieve can influence data needs. As such, experimentation on a case-by-case basis is crucial to determine whether the labelled data at hand should be regarded as considerable or limited. In our experience, and for the sake of offering ballpark figures, having fewer than 100 data points tends to constitute a limited amount. A considerable amount of data, on the other hand, is likely to materialize within the range of 500 to 5000 labelled data points.

We need to highlight three important aspects related to the process in Figure~\ref{fig:decision}. First, no individual decision model can encompass the full spectrum of techniques employed in NLP4RE. Our model aims to offer a simplified representation of common scenarios, rather than imposing constraints or promoting a lack of flexibility in technique selection. Second, the process is likely to evolve in the future to stay in step with NLP4RE research. In particular, capitalizing on the interactive capabilities of large language models, there is potential to further elaborate the process by considering prompting strategies and providing additional guidelines. However, due to the scarcity of NLP4RE approaches built on large language models, we have to defer doing so to the future. Finally, when selecting enabling techniques for analysis, the cost and environmental impact must be considered. Most notably, the resource-intensive nature of large language models requires justification, especially when alternatives cannot be dismissed due to compelling reasons such as lack of accuracy.

\subsection{Post-processing}\label{subsec:post}
Post-processing -- the third step in the process of Figure~\ref{fig:overview} -- aims to enhance the results of the Analysis step or to adapt these results for human analysts' better understanding. Post-processing is not needed in all NLP4RE solutions and is thus an optional step.

To illustrate a simple scenario where post-processing is required, let us consider the task of requirements identification. For this task, one may apply the following heuristic as a post-processing step: if all but one sentence in a passage are categorized as requirements (during the Analysis step), that lone sentence should be reclassified from a non-requirement to a requirement. This adjustment will likely increase the accuracy of requirements identification~\cite{Abualhaija:20b}.

For a more advanced example of post-processing, let us consider requirements completion based on predictions by a language model. In this context, the language model is likely to generate a non-negligible number of predictions that are not useful (false positives) alongside the useful ones. To reduce the incidence of unuseful predictions in the final results, the development of a post-processing filter becomes essential~\cite{Luitel:23}.

In its simplest form, post-processing can be light, e.g., in the case of the heuristic mentioned earlier for requirements identification. In more complex scenarios, like the one mentioned above where predictions need to be filtered, additional enabling techniques might be needed to carry out post-processing.
\section{Enabling Techniques: Overview and Guidelines}~\label{sec:enablingTechniques}

In this section, we outline the various enabling techniques shown in Figure~\ref{fig:overview} and provide practical guidelines for applying and evaluating them.

\subsection{NLP Pipeline}~\label{subsec:nlp}
The NLP pipeline is a sequence of modules that incrementally add linguistic annotations to an input text. The pipeline typically begins with tasks like tokenization (breaking text into words or sub-words), sentence splitting (segmenting a passage into sentences), and lemmatization (reducing words to their base forms). Next and depending on the annotations required, the pipeline performs tasks such as part-of-speech (POS) tagging (labelling words with their grammatical roles), named-entity recognition (identifying entities like names, dates, and locations), and syntactic parsing (analyzing sentence structure). Syntactic parsing includes two main techniques: constituency parsing (deconstructing sentences into grammatical constituents) and dependency parsing (determining grammatical relationships between words). 

\begin{figure}
    \centering
\includegraphics[width=\textwidth]{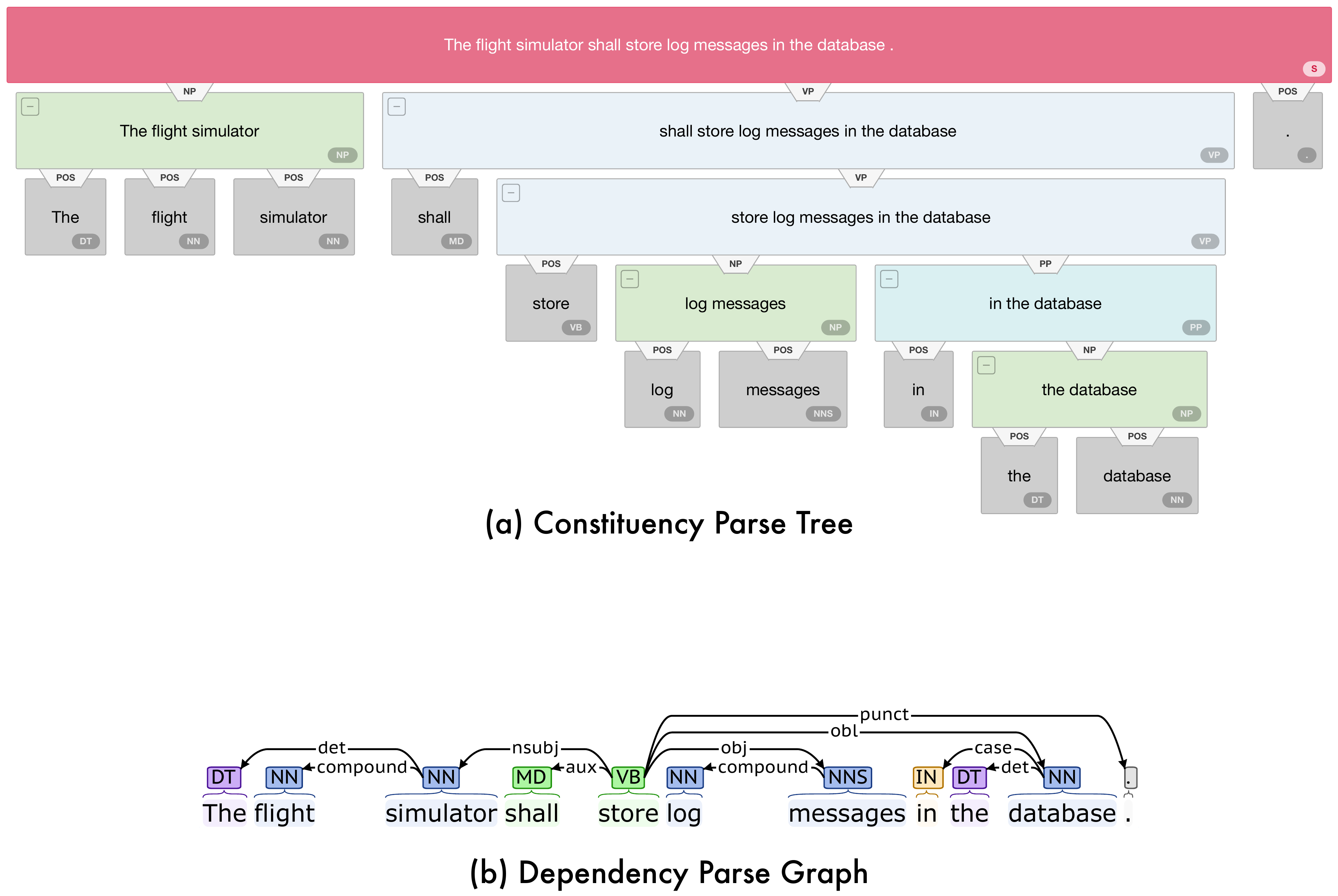}
    \caption{Illustration of (a) Constituency Parsing and (b) Dependency Parsing. Both Parsing Methods Require Sentence Detection and POS Tagging.}
    \label{fig:nlp}
\end{figure}

Figures~\ref{fig:nlp} (a) and (b) respectively exemplify the annotations generated by the constituency parsing and dependency parsing modules for the following requirements sentence: $R=\emph{``The flight simulator shall store log messages in the database.''}$. 
The annotations generated by the constituency parser for this sentence are: NP (noun phrase), VP (verb phrase), and PP (prepositional phrase). These annotations capture the hierarchical constituents of the sentence. The relationships between the words in the sentence are given by the pairwise links generated through dependency parsing. For instance, from the dependency parse graph of Figure~\ref{fig:nlp}~(b), one can determine that ``simulator'' functions as the subject (nsubj) for ``store,'' and that ``messages'' serves as the object (obj) for this transitive verb. Crucially, both parsing methods require sentence splitting and POS tagging. Figure~\ref{fig:nlp}(a) illustrates sentence annotation (S), while both figures include POS tags (DT: determiner, NN: noun, NNS: plural noun, MD: modifier, VB: verb, IN: preposition).

When using the NLP pipeline, it is important to consider the specific natural language(s) that require support. 
While most NLP modules have good performance over well-written English, their effectiveness can vary significantly when dealing with other languages or when processing text that deviates from grammatical norms, such as user feedback (e.g., app reviews) or developer commit messages. Experimentation is therefore typically necessary for constructing an accurate NLP pipeline. NLP workbenches such as GATE~\cite{GATE} and DKPro~\cite{DKPro} facilitate the integration of NLP modules from different NLP libraries. For instance, these workbenches enable the use of the POS tagger provided by one library, say, Apache OpenNLP~\cite{OpenNLP}, alongside the constituency and dependency parsers from another library, say, Stanford CoreNLP~\cite{CoreNLP}. This flexibility to combine NLP modules from different libraries enables systematic experimentation with various pipeline configurations to determine the optimal configuration for the task at hand. An example of such experimentation can be found in our work on checking conformance with requirements templates~\cite{Arora:15a}.

If systematic experimentation with alternative NLP modules is not feasible or if there are constraints on using a single library, e.g., to minimize complexity, it would be important to conduct an error analysis on the annotations generated by the NLP pipeline. This analysis helps with ensuring the absence of systemic issues. Common systemic issues include recurring errors in sentence detection, repeated inaccuracies in tokenization and POS tags, and incorrect parsing results.

\begin{takeaway*}{NLP Pipeline}
 Prioritize pipeline implementations that are known to work well for the language(s) you need to support. Experimentation with different NLP modules is often useful for improving accuracy. If extensive experimentation with the NLP pipeline is not possible or you are limited to a single library, conduct an error analysis on the annotations to identify and address any systemic issues such as recurring errors in sentence detection, POS tags, and parsing.
\end{takeaway*}
\subsection{Relevance Measures}~\label{subsec:measures}
In the context of RE, \emph{relevance} denotes the extent to which various requirement segments are related or how closely a specific requirements segment aligns with a particular query, topic, or artifact (e.g., a design document or a portion thereof). The concept of relevance is typically quantified using one or a combination of three distinct categories of metrics:  \emph{syntactic}, \emph{semantic}, and \emph{statistical}. 
\begin{itemize}
\item\sectopic{Syntactic Relevance} is the string-based or structural relatedness between two or more text segments. There are numerous syntactic relevance measures.  For a fairly comprehensive list of syntactic measures, consult~\cite{Gomaa:13}. An example metric in the syntactic category is Levenshtein~\cite{Gomaa:13}. This metric calculates the minimum number of single-character edits (insertions, deletions, or substitutions) required to transform one string into another. Levenshtein distance is usually normalized (scaled from 0 to 1) by dividing it by the maximum possible edit distance between the two strings. For instance, the (normalized) Levenshtein distance between the phrases ``software system'' and ``software systems'' is 0.9375, indicating that the two strings have considerable lexical overlap. 

\item\sectopic{Semantic Relevance} goes beyond string matching to measure similarity between word meanings.  Semantic relevance is typically quantified using metrics like PATH within a semantic network  such as WordNet~\cite{wordnetOnline}.   
WordNet captures various relationships, such as hypernymy and hyponymy, which denote an ``is-a'' connection, and meronymy and holonymy, representing a ``part-of'' association. To illustrate, consider the relationship between ``vehicle'' and ``car'', where ``vehicle'' serves as the hypernym of ``car'', and the relationship between ``wheel'' and ``car'', with ``wheel'' acting as the meronym of ``car''. PATH similarity calculates the  shortest path between two concepts on the ``is-a'' hierarchy. For instance, the PATH score between the terms ``cat'' and ``mammal'' is higher than between ``cat'' and ``vehicle'', indicating that ``cat'' is semantically more related to ``mammal'' than ``vehicle''.

\item\sectopic{Statistical Relevance} assesses relevance by analyzing the frequency and distribution of terms within a document, often employing algorithms such as TF-IDF and BM25. These algorithms are frequently normalized on a scale from 0 to 1~\cite{Baeza:99}. Statistical methods typically operate at the level of term frequency and do not necessitate a pre-constructed semantic network. For instance, TF-IDF evaluates a term's importance within a document based on its frequency of occurrence in that document, normalized by its frequency across the entire corpus. BM25, which extends TF-IDF, takes into account further factors like term saturation and document length.
To illustrate, consider a corpus of requirements documents. A term like ``user authentication'' might appear infrequently but could be highly significant. BM25 can rank a document that extensively discusses this term higher than a document that only mentions it in passing, thereby indicating its greater relevance in the context of  security requirements. Statistical measures are more commonly employed in information retrieval (IR) problems within NLP4RE, such as querying requirements~\cite{Abualhaija:22}.
\end{itemize}

\sectopic{Applications and Other Considerations.} Relevance measures serve two main use cases in NLP4RE. The first is to calculate similarity either within a set of requirements or between requirements and textual segments found in other development artifacts. The second use case is to assess the relative significance of terms in documents. Frequently, relevance measures are employed as features in both supervised and unsupervised learning.

Different relevance measures can be combined to provide a more holistic characterization of relevance. For example, when tasked with constructing a requirements glossary, the combination of syntactic and semantic measures can help identify a wider range of variations among related domain terms. To illustrate, if our objective is to cluster terms related to a flight simulation system, we anticipate that ``flight coordinates'', ``aircraft position'', and all concepts associated with ``flight positioning'' should land in the same cluster~\cite{Arora:17}. Simultaneously applying both syntactic and semantic measures facilitates the determination of these terms being highly similar.

In relation to syntactic measures, it is important to note that, because requirements frequently manifest variability in phrasing, vocabulary selection and syntactic structure, techniques like lemmatization and tokenization are often required prior to computing syntactic measures. This preprocessing helps mitigate variability and enhances the accuracy of comparisons~\cite{Arora:17}.

Finally, and in relation to semantic measures, we note that these measures are quickly being replaced by more advanced techniques, notably embeddings. Embeddings not only capture semantic similarity but also contextual similarity, as we discuss in Section~\ref{subsec:embeddings}. Furthermore, while lexical resources like WordNet offer the advantage of establishing human-interpretable connections between words, techniques such as embeddings provide a more nuanced characterization  of meaning, although they may not be entirely interpretable by humans. Consequently, when the primary goal is the application of similarity metrics, rather than explaining relationships, there is often limited justification for employing semantic measures like PATH in future research.

\begin{takeaway*}{Relevance Measures}
  Combining different relevance measures often results in more accurate analytical outcomes. Relevance measures can be applied at different levels of granularity, ranging from individual tokens (e.g., Levenshtein distance) to entire documents (e.g., TF-IDF). Empirical evaluations of relevance measures can centre around identifying the most effective combination of these measures or benchmarking advanced solutions against relevance measures considered as baseline methods.
\end{takeaway*}
\subsection{Embeddings}~\label{subsec:embeddings}
Embeddings enable the conversion of words into numerical vectors. These vectors encapsulate the semantic connections among words, in turn supporting a more meaningful manipulation of language. Word embeddings are typically derived through self-supervised approaches such as Word2Vec~\cite{Mikolov:13c}, GloVe~\cite{Pennington:14}, and the pre-training of language models such as BERT~\cite{Devlin:18} and GPT~\cite{Radford:18}. For example, using the 300-dimensional variant of GloVe embedding vectors, the word ``requirements'' would be represented as a 300-dimensional vector:  
$[\text{-1.3598e-01}, \text{-1.8174e-01},\cdots, \text{-6.2015e-02}]$.

While the primary goal of embeddings is to represent individual words, methods also exist for generating embeddings for phrases and sentences. For example, a simple approach for obtaining sentence embeddings is to compute the weighted average of the word embeddings in a given sentence~\cite{Arora:17_Embeddings}.

There are two main use cases for embeddings in the existing NLP4RE literature: (1)~computing semantic similarity, typically through the cosine measure, and (2) using embeddings as features for  learning. To illustrate, suppose that we are interested in identifying most similar requirements, e.g., as a way to find overlapping or redundant requirements. Consider the following three statements: 
\textsf{R1} = \emph{``The system shall react to user input within one second.''}; \textsf{R2} = \emph{``The system shall respond within one second.''}; and  \textsf{R3} = \emph{``The system shall encrypt sensitive data.''}. For a requirement sentence $R$, let $\emph{emb}(R)$ denote the sentence's embeddings. By utilizing the 300-dimensional variant of GloVe and employing averaging to derive sentence embeddings from word embeddings, we would obtain the following: $\textit{cosine}(\emph{emb}(\textsf{R1}), \emph{emb}(\textsf{R2})) \approx 0.95 > \textit{cosine}(\emph{emb}(\textsf{R1}), \emph{emb}(\textsf{R3})) \approx 0.83 > \textit{cosine}(\emph{emb}(\textsf{R2}), \emph{emb}(\textsf{R3})) \approx 0.78$. Now, the requirements analyst can, for example, sort the requirements pairs in descending order of similarity and inspect the most similar pairs to determine if there are overlaps or redundancies. Alternatively, when it is feasible to create a labelled dataset for training, the embeddings can be used as features -- either on their own or alongside other features -- for building a feature-based classifier that distinguishes similar and non-similar requirements pairs. In our illustrative example, and assuming that only \texttt{R1} and {R2} are deemed similar, one could infer (among other labelled data points) the following for training: $\emph{emb}(\textsf{R1})|\emph{emb}(\textsf{R2})|\texttt{SIMILAR}$ and $\emph{emb}(\textsf{R1})|\emph{emb}(\textsf{R3})|\texttt{NOT-SIMILAR}$. Here, ``$|$'' denotes vector concatenation, and ``\texttt{SIMILAR}'' and ``\texttt{NOT-SIMILAR}'' denote labels for pairs of requirements.


There are some important considerations to note when working with embeddings:

\sectopic{Dimensionality of Embeddings.}
The dimensionality of embeddings, which refers to the number of dimensions (or features) used to represent each word as a vector, determines the richness of information in word vectors. While some dimensions may have clear interpretations, e.g., gender or sentiment, others can be complex, capturing subtle and abstract aspects of word semantics that would be difficult for humans to interpret directly. There is no universally suitable choice for the dimensionality of word embeddings. Although a larger dimensionality can provide increased semantic nuance and potentially improved accuracy, one must consider the curse of dimensionality~\cite{Mikolov:13c} -- the inherent challenges that arise when dealing with high-dimensional data. We recommend experimentation with alternative embedding methods and dimensionality options to find an acceptable trade-off between accuracy for the analytical task at hand and the challenges posed by high-dimensional data. Note that the techniques discussed in Section~\ref{subsec:ML} for feature selection and reduction can also be applied to  embeddings to reduce dimensionality.

\sectopic{Non-contextual vs. Contextual Embeddings.}
Embeddings can be either non-contextual or contextual. Non-contextual embeddings, such as those produced by GloVe, create fixed word vectors that remain the same regardless of context. In contrast, contextual word embeddings, as produced by models like BERT and GPT, consider the surrounding words to generate word vectors that adapt to the context. To illustrate, consider the following two sentences: \textsf{S} = \textit{``Meeting privacy requirements is essential.''}; and $\textsf{S}'$ = \textit{``The system shall meet all the privacy requirements stipulated by the GDPR.''} Using GloVe to obtain embeddings for the word ``requirements'' yields the same vector for both sentences. In contrast, if one uses models such as BERT or GPT, the embeddings obtained for the same word in different sentences would differ. For instance, the \textsf{bert-base-uncased} variant of BERT yields the following vectors for ``requirements'' in \textsf{S} and $\textsf{S}'$, respectively: $[\text{2.7466e-01},  \text{4.8193e-01}, \cdots, \text{4.1681e-01}]$ and $[\text{-1.6852e-01,  3.5143e-01}, \cdots,  \text{3.5699e-01}]$. This difference is due to the contextually different meaning of ``requirements'' in these two sentences, where the word conveys the sense of a prerequisite condition in \textsf{S} and the sense of a specification in $\textsf{S}'$.

Contextual embeddings offer greater accuracy but come at a higher computational cost. It is therefore important not to dismiss non-contextual embeddings outright but to weigh their potential alongside contextual embeddings to determine whether the added accuracy of contextual embeddings is worth the extra cost.

\sectopic{Domain-specific Embeddings.}
Domain-specific embeddings capture context-specific word meanings that generic embeddings may miss. For instance, specialized BERT variants, such as BioBERT~\cite{Lee:20} for biomedical texts and LegalBERT~\cite{Chalkidis:20} for legal documents, provide domain-specific embeddings for their respective domains. When domain-specific embeddings exist, it is worthwhile to compare them with generic embeddings for potential improvements. 
Further, in cases where domain-specific embeddings are unavailable, but a suitable domain-specific corpus is accessible, one can attempt to construct domain-specific embeddings from scratch. Recent efforts in software engineering, such as building Word2Vec and GloVe embeddings for model-driven engineering~\cite{Lopez:23}, provide useful guidance in this regard.

\vspace*{1em}

\begin{takeaway*}{Embeddings}
 Experiment with different embedding technologies and dimensionality options to strike a balance between accuracy and the challenges of high-dimensional data. Assess whether the added accuracy of contextual embeddings justifies their higher computational cost, or if non-contextual embeddings suffice for your specific task. Explore domain-specific embeddings when available, as they may be superior at capturing context-specific meanings.
\end{takeaway*}
\subsection{Query-based Pattern Matching}~\label{subsec:Rules}
The annotations produced by the NLP pipeline provide a rich basis for defining queries that detect patterns of interest within an input text. Typically, pattern-matching queries work with ``spans''. Each span represents a distinct sequence of consecutive words or tokens within the given text.

An important technical aspect in pattern matching is the choice of the query language. In NLP, span information can be flat, focusing on individual words and their properties (like POS tags), hierarchical, revealing how words combine into larger structures (as in constituency parsing), or graph-based, capturing relationships between words (as in dependency parsing). Different NLP toolkits offer different query languages; CoreNLP~\cite{CoreNLP}, for instance, provides TokensRegex for token-based regular expressions, Tregex for tree-based linguistic structures, and Semgrex for syntactic dependency patterns. To illustrate, we exemplify these query languages:
\begin{enumerate} 
\item[(a)] The  TokensRegex query \textsf{[\{tag:/VB.*/\}]} extracts all spans tagged as verbs.

\item[(b)] The Tregex expression \textsf{NP[$<$NN $\mid$ $<$NNS]} extracts all spans that are Noun Phrases (NPs) and immediately dominate a singular noun (NN) or a plural noun (NNS).

\item[(c)] The Semgrex query \textsf{\{pos:/VB.*/\} $>$nsubj \{\}=subject $>$obj \{\}=object} extracts verbs that both have a subject and an object, alongside the subject and the object.
\end{enumerate}

\begin{figure}
    \centering
\includegraphics[width=\textwidth]{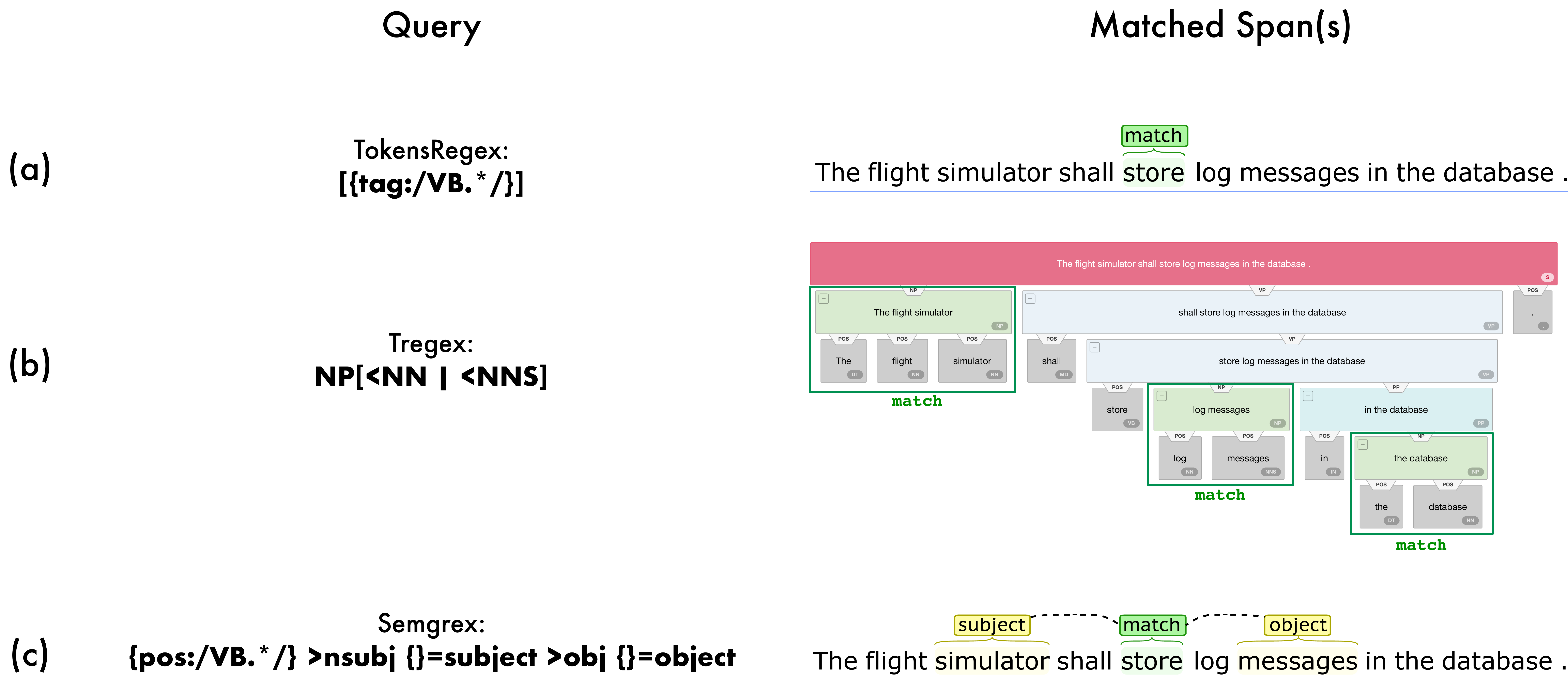}
    \caption{Illustration of Query-based Pattern Matching over the Annotations of Fig.~\ref{fig:nlp}.}
    \label{fig:patterns}
\end{figure}

Figure~\ref{fig:patterns} shows the results of the above queries as applied to the annotations in Fig.~\ref{fig:nlp}. All three queries serve a purpose in NLP4RE. Query~(a) produces results that can be used as a feature for identifying requirements, grounded in the hypothesis that a higher verb count signifies a higher likelihood of requirements being present. Query~(b) identifies constituent noun phrases and verb phrases within a requirements statement. This information is valuable for various purposes, one of which is to validate whether the statement conforms to a specific template, 
such as EARS~\cite{Mavin:12}. Query (c) is useful for constructing a domain model~\cite{Arora:16}; the query extracts a probable association. In the case of our example, the association would be ``[flight] simulator \emph{stores} [log] messages''.

Query-based pattern matching techniques often face criticism related to both the scope of the study that informs query development and the applicability of the resulting queries beyond their initial purpose. To address these concerns, we recommend a two-fold approach. First, it is important for query designers to carefully document the query specification process, providing a clear explanation of the considered documents and domains. The primary objective should be to enhance empirical reliability, ensuring that the process is as reproducible as possible by others. Second and concerning generalizability, one should avoid using all the text under analysis for query development. This approach allows for the assessment of query generalizability on unseen text, resulting in a more credible evaluation. An even more robust strategy involves expanding the analysis beyond the immediate domain under study. For example, if the focus is on regulatory documents, it is common to derive queries in one legal jurisdiction, such as Europe, and subsequently evaluate query accuracy in texts from another jurisdiction, like Canada~\cite{Sleimi:21}. By adhering to these basic principles — detailed documentation of the query derivation process and assessments of generalizability — one can enhance the utility of query-based pattern matching techniques.

\vspace*{1em}

\begin{takeaway*}{Query-based Pattern Matching}
(1) When working with NLP annotations, select query language(s) that align with your specific analysis needs. (2) To ensure the reliability of query derivation, carefully document the steps you take to create the queries. This documentation should include the criteria, context, and content used for query design, making it easier for others to reproduce your work. (3) Refrain from using all the text you have for query development. To improve the applicability of your queries, evaluate their performance on unseen (or withheld) text or in different domains.
\end{takeaway*}
\subsection{Feature-based Machine Learning}~\label{subsec:ML}
Feature-based machine learning (ML) is concerned with using labelled data to train algorithms to make predictions based on predefined features. Feature-based ML has numerous applications in NLP4RE. For instance, it can be used for (1) categorizing input requirements as either functional or non-functional~\cite{Kurtanovic:17}, or (2) predicting the number of story points for user stories based on historical data~\cite{Choetkiertikul:19}. The former application is a \emph{classification} problem, entailing the assignment of discrete labels to data points (``functional'' or ``non-functional''); whereas the latter scenario constitutes a \emph{regression} problem as it involves estimating numeric values (story points).

Building a feature-based ML model has five main steps. Below, we discuss these steps and outline some important practical considerations related to each step.
\begin{enumerate}
    \item\emph{Data collection} involves gathering relevant and sufficient data, including the labelling process. This phase can be very time- and effort-intensive, particularly when manual labelling is required. To ensure consistency, it is crucial to define a clear and agreed-upon labelling task with protocols that annotators can consistently follow. To enhance internal validity, authors should avoid annotating their test data. When feasible, it is further important to employ multiple annotators with some overlap to enable the calculation of inter-annotator agreement. Detailed documentation of the data-collection protocol is essential for reliability~\cite{Abualhaija:20b, Abualhaija:22}.   
    
\item \emph{Data wrangling} is concerned with handling missing data, noise removal (e.g., stop words or redundant information), deciding about how to manage structured and unstructured data, and transforming results into the desired format. It is crucial to thoroughly document all techniques and decisions made in this step, as even minor changes can significantly impact the accuracy of the ML model~\cite{Fan:23}.

\item \emph{Feature engineering} is the process of defining relevant features for learning from the input data. In NLP4RE, features are usually defined over the outputs of the pre-processing techniques discussed in Sections~\ref{subsec:nlp}--\ref{subsec:embeddings}. For instance, a feature based on the NLP pipeline could be the number of verbs in a sentence. A feature using relevance measures might be the highest TF-IDF rank of a specific term across all articles in a corpus. Finally and in relation to embeddings,
the components within an embedding vector can be regarded as features, as elaborated in Section~\ref{subsec:embeddings}.

Features should be treated as hypotheses for addressing the problem at hand, and their usefulness must be evaluated. Techniques for computing feature importance, e.g., information gain~\cite{Azhagusundari:13}, and PCA analysis~\cite{Abdi:10}, may be used for eliminating redundant or less significant features, thereby preventing overfitting. Feature engineering must also prioritize computational efficiency, particularly in real-time tasks such as live editing of requirements~\cite{Abualhaija:20b}, where almost instant feature computation is necessary. In such scenarios, certain computationally expensive features may need to be sacrificed, even if they offer marginal improvements. For example, the resource-intensive process of constituent parsing, despite its potential value, may prove impractical. Conducting a cost-effectiveness analysis, weighing the cost of computing each feature against its impact on the resulting ML model's accuracy is therefore recommended.
 
\item \emph{Model Selection, Tuning, and Training} is concerned with choosing an appropriate ML algorithm, fine-tuning its hyperparameters, and training it over feature-engineered data~\cite{Feurer:19}. Initially, the computed feature data should be split into three sets: the training set (typically 70\% of all the data), the validation set (usually 10\% of the data) for model selection and hyperparameter optimization, and the test set (typically 20\% of the data) to evaluate the trained ML model~\cite{Raschka:18}. During dataset partitioning, it is paramount to avoid data leakage, which means avoiding any exposure of test set data during training and validation~\cite{Shabtai:12}. If the dataset covers multiple projects, it is advisable to ensure that the test set includes data from `hold-out' projects not involved in training or validation at all~\cite{Raschka:18}.

NLP4RE frequently encounters a scarcity of the kind of ``abundant'' data necessary for training complex ML algorithms, such as deep learning. In such instances, ML algorithms like Random Forest can provide scalable solutions, even when dealing with smaller datasets, noting of course that this advantage comes at the cost of requiring manual feature engineering.

Model tuning is another vital step in the process. While hyperparameter optimization should theoretically be integrated with model selection, the number of combinations to evaluate when combining model selection and hyperparameter optimization can be exceedingly large~\cite{Feurer:19}. For example, systematically exploring discrete values across key hyperparameters in the Random Forest algorithm can result in over 500 combinations. Given this challenge, a more practical approach would be to initially select a suitable ML algorithm by experimenting with multiple algorithms at their default settings, and then proceed to fine-tune the hyperparameters for the chosen algorithm. The choice of hyperparameter tuning strategy depends on available resources, such as an extensive grid search or a lightweight stochastic random search, for example.

\item \emph{Evaluation} involves assessing a trained ML model's performance on a test set using metrics such as accuracy, precision, recall, F-score, and mean absolute error (MAE). Depending on the context, one can opt for ``hold-out'' test sets or k-fold cross-validation. The selection and prioritization of metrics require special attention. For instance, in a binary classification task where 95\% of samples belong to Class A and only 5\% to Class B, a classifier exclusively predicting Class A would achieve 95\% accuracy. Consequently, accuracy may not be a suitable metric when the dataset is imbalanced. Another notable point related to metrics is that, in NLP4RE, recall often outweighs precision in terms of importance, typically making recall a priority in solution development~\cite{Berry:21}. In relation to reporting, care needs to be taken when reporting aggregate metrics like F-score: such metrics should be reported alongside the source metrics (in this case, precision and recall) rather than as substitutes. 

Beyond reporting metrics, it is important to reflect on how these metrics translate into either benefits or drawbacks for users. For example, it is always valuable to contemplate the significance of various types of classification errors and determine their impact on users. When the cost of prediction errors is deemed too high, e.g., when the ML model's predictions serve as recommendations requiring meticulous manual follow-up, one may consider implementing a human feedback loop to continuously retrain the model and reduce errors~\cite{Arora:19}.
\end{enumerate}

\begin{takeaway*}{Feature-based ML}
 Develop explicit procedures for manual data labelling during data collection and document these procedures. Keep track of data-wrangling decisions, as they can have a substantial impact on the outcomes.  Approach features as hypotheses and confirm their significance before incorporating them into the final solution. Given the numerous combinations available for hyperparameter tuning during model selection and tuning, you may want to begin with experimentation using the default settings of machine learning algorithms.
Ensure there is no data leakage, which means avoiding the inadvertent exposure of parts of test documents during the training and validation processes. The choice of evaluation metrics is of great importance and should be aligned with the specific NLP4RE task at hand.
\end{takeaway*}
\subsection{Clustering Algorithms}~\label{subsec:clustering}
Clustering algorithms group similar data points into subsets or clusters to reveal patterns and structures within the data. This is achieved using a quantitative measure of  similarity and ensuring that points in the same cluster are more similar to each other than to those in different clusters.
In NLP4RE, relevance measures (Section~\ref{subsec:measures}) and embeddings (Section~\ref{subsec:embeddings}) 
are commonly used to compute similarity for clustering purposes.  For instance, using GloVe embeddings, requirements statements can be transformed into vectors and then clustered using clustering algorithms such as K-means, agglomerative clustring or expectation maximization (EM)~\cite{Witten:16}. To illustrate, consider a system with six requirements: \textsf{R1-R3} from Section~\ref{subsec:embeddings} and \textsf{R4-R6} defined as follows: \textsf{R4} = \emph{``The system shall allow users to customize the UI theme, available as `dark' and `light' versions.''}; \textsf{R5} = \emph{``The system shall allow users to sync data across multiple devices.''}; and \textsf{R6} = \emph{``The system shall allow users to reset  passwords only after email authentication.''}. Once sentence embeddings have been generated,  cosine similarity between requirements pairs can be used as the basis for clustering. An illustrative clustering of these requirements, aimed at determining implementation responsibilities, might consist of $[\textsf{R1}, \textsf{R2}]$ (system responsiveness), $[\textsf{R3}, \textsf{R6}]$ (system protection), $[\textsf{R4}]$ (system UI customization), and $[\textsf{R5}]$ (system data management). Clustering has numerous applications in NLP4RE, including tasks such as traceability link retrieval, identifying overlapping or redundant requirements, and categorizing app reviews to pinpoint new feature requests. Below, we present some important practical considerations for an efficient utilization of clustering algorithms in NLP4RE.

\sectopic{Determining the Number of Clusters (k).}
Choosing an appropriate  number of clusters ($k$) is an important prerequisite  for effectively applying many common clustering algorithms such as K-means and EM. Techniques like the Elbow method, Silhouette analysis, and Bayesian Information Criterion (BIC) can help estimate an appropriate $k$~\cite{Witten:16}. Another alternative is a recent summarization metric proposed in clustering app reviews~\cite{Nema:22}. Domain knowledge remains a key factor alongside these methods in deciding the number of clusters. The appropriate value of $k$ is influenced by the task at hand and how users plan to utilize the resulting clusters. 
For instance, when clustering requirements terms, one may emphasize creating numerous small clusters, e.g., to simplify glossary construction~\cite{Arora:17}. On the other hand, when clustering core requirements within software product lines, it may be preferable to have a smaller number of clusters, as this streamlines the reviewing of common functions~\cite{Reinhartz:20}.

\sectopic{Hierarchical vs. Partitional vs. Soft Clustering.} In partitional clustering algorithms, such as K-means, data is divided into non-overlapping clusters without any inherent structure. In contrast, hierarchical clustering creates a tree-like structure. Partitional methods are more suitable when the primary focus of the analysis is to rapidly identify overarching themes. For instance, Di Sorbo et al.~\cite{DiSorbo:16} adopt a partitional approach to categorize app reviews into pre-defined topics, determining the necessary changes in the apps according to user feedback. Hierarchical clustering, on the other hand, is better suited when there is a need to comprehend and visualize the relationships and nested structures within the dataset. For instance, Reinhartz-Berger and Kemelman~\cite{Reinhartz:20} use hierarchical clustering to explore the relationships between software product line requirements from different products by clustering the core requirements. It is worth noting that hierarchical clustering algorithms also provide mechanisms to create partitions. In soft clustering (also know as fuzzy clustering), each data point can belong to multiple clusters with varying degrees of membership. Soft clustering is ideal for situations where the boundaries between clusters are not rigidly defined, and instances may possess properties of multiple clusters. An example of this approach is the EM algorithm. An RE-related use case is clustering of requirements terms. In this context, it is beneficial to allow each term to have membership in different clusters~\cite{Arora:17}. For instance, the term ``flight simulator'' may find membership in both the ``flight''-related and ``simulator''-related clusters within a requirements document. Ultimately, the choice between partitional, hierarchical, or soft clustering depends on the specific task and analysis objectives at hand. Analysts should therefore  understand the distinctions and potential applications of different types of clustering algorithms to ensure that their chosen algorithm aligns with the goals of their analysis.

\sectopic{Evaluation.}
Cluster evaluation can be  either \emph{internal} or \emph{external}. Internal evaluation assesses the quality of clusters without relying on external labels, focusing on the principles of \emph{cohesion} and \emph{separation}. Cohesion measures how closely related the members of the same cluster are, reflecting the compactness of the clusters, while separation assesses how distinct or well-separated a cluster is from others. A higher separation implies that clustering can more effectively distinguish between clusters.
External evaluation, on the other hand, measures the quality of clusters by comparing them to a  ground truth. This comparison can involve evaluating the overlap or mutual information between the original clusters and the generated ones. Developing a ground truth in clustering is challenging due to the inherent subjectivity of this task and its context dependence. Requirements and related artifacts can have multifaceted interpretations, leading to multiple valid ways of clustering based on different perspectives or objectives. Creating a good ground truth necessitates: (i) a clear understanding of the concept of clustering as well as the end goal to achieve, (ii) substantial time and effort, and (iii) vigilance against bias and inconsistency, as different experts may have contrasting opinions based on their experiences, potentially resulting in inconsistencies and  posing threats to both internal and construct validity.

\vspace*{1em}

\begin{takeaway*}{Clustering Algorithms}
 Important decisions in clustering include determining a suitable number of clusters (e.g., the Elbow method or Silhouette analysis), and opting between clustering types (partitional, hierarchical, or soft). The choice of clustering algorithm depends on the goal of the analysis and requires domain understanding. Evaluating the effectiveness of clustering is critical, which can be approached internally by assessing the cohesion and separation of clusters or externally by comparing against a ground truth. An external evaluation often poses challenges, as creating a ground truth for clustering is subjective and often requires substantial effort.
\end{takeaway*}
\subsection{Large Language Models}~\label{subsec:LLMs}
Language models (LMs) are statistical models that use neural networks to predict the next word in a sequence based on preceding words. For example, given the text ``Paris is the capital of'',  a langauge model may predict the next word as ``France''. Language models increasingly serve as the foundation for various downstream NLP tasks such as text completion, translation, and summarization. Large language models (LLMs) like GPT-4 and BERT are scaled-up versions of the LM concept, with billions and sometimes even trillions of parameters. 
The descriptor ``large'' in LLM pertains to the model's size in terms of the number of parameters. These parameters represent the neural network layers' weights that the model acquires through training.  Generally, larger models tend to perform better at comprehending context, drawing inferences and producing answers that resemble human responses.

\sectopic{Hyperparameters.} LLMs have a range of hyperparameters that can be adjusted during both training and fine-tuning. The hyperparameter profiles can vary among different LLMs, and the extent to which individual LLMs allow end-users to modify these hyperparameters also varies.
Some common LLM hyperparameters include:
(i)~\textit{learning rate}, determining how quickly the model adapts its weights during training;
(ii)~\textit{number of epochs}, referring to how many times the model will go through the entire training dataset; 
(iii)~\textit{batch size}, denoting the number of training examples utilized in one iteration;
(iv)~\textit{sequence length}, referring to the number of tokens that the model reads in one go;
(v)~\textit{dropout rate}, referring to the fraction of input units to drop during training to mitigate overfitting; and (vi)~\textit{temperature}, a parameter that modulates the probability distribution over the predicted words, making the outputs more focused (lower values) or more random (higher values).

Extensive experimentation across a broad range of LLM hyperparameter combinations currently presents challenges due to constraints in both time and budget. Nevertheless, it remains essential to select a practical subset of hyperparameter combinations that can be explored within the available resources. This approach enables a more informed understanding of how hyperparameters influence the performance of an LLM in conducting a specific task.

\sectopic{Prompting.} Prompting refers to providing a specific input or query to guide a generative AI model in producing the desired output, such as text or images~\cite{Hariri:23}. Prompts, which are concise textual inputs given to generative AI models, including LLMs, convey information about the specific task that the model is expected to execute. 

Creating prompts that are effective in eliciting the desired output from an LLM requires good prompt engineering. Prompt engineering involves the selection of appropriate prompt patterns and prompting techniques~\cite{White:23}. Prompt patterns encompass various templates tailored for specific objectives. For example, the output customization pattern focuses on refining the format or structure of LLM-generated output, with the LLM often adopting a particular persona (role) while generating the output. Prompting techniques, on the other hand, are strategies to extract the best possible output from LLMs. Example prompting techniques include zero-shot prompting, few-shot prompting, chain-of-thought prompting, and tree-of-thought prompting~\cite{Liu:23}.

When conducting empirical examinations of LLM-based solutions, it is important to take into account the alternative choices that one can make during prompt engineering. These alternatives should ideally be compared through empirical means. Nonetheless, much like the challenges encountered when exploring hyperparameters, the vast array of possible combinations of prompt patterns and prompting strategies may be too numerous to thoroughly investigate. Another important aspect to consider is that even minor alterations in prompts can lead to considerable variation in the outputs generated by LLMs. In view of this, empirical studies should also assess prompt robustness by examining multiple variants of the same prompt.

\sectopic{Fine-tuning.} 
While LLMs come pre-trained on very large corpora, they often require some level of fine-tuning to specialize them for particular tasks or domains. When the requirements automation task at hand aligns closely with common knowledge, such as ambiguity handling, one might achieve good results with little or no fine-tuning. However, if the task is specialized and involves RE-specific semantics, such as requirements classifications, fine-tuning often becomes necessary to ensure accurate results.
When fine-tuning an LLM, one needs to be cognizant of the non-deterministic nature of the process, which is influenced by factors such as random initialization and regularization. Depending on the extent and diversity of the fine-tuning data, significant variations may be observed in results across different fine-tuning attempts. To account for this randomness, we recommend fine-tuning LLMs multiple times and reporting average results rather than relying on the outcome of a \hbox{single fine-tuned model.}

\begin{takeaway*}{Large Language Models}
  When working with LLMs, explore hyperparameters within resource constraints. 
  Invest in prompt engineering, using suitable prompt patterns and techniques, and empirically compare different alternatives. Take note of the non-deterministic nature of fine-tuning, and present results averaged over multiple fine-tuning runs.
\end{takeaway*}

\section{Summary and Conclusion}~\label{sec:conclusion}
The main goal of this chapter was to aid newcomers in the selection of appropriate NLP4RE techniques and the application of some essential principles for their evaluation. We acknowledge the wide array of NLP technologies employed in RE. Not all NLP4RE approaches may neatly align with the general, and sometimes simplified, framework we have outlined in this chapter. 

This chapter was written during a transformative period in the NLP field, spurred by the emergence of generative AI and large language models. We hope that this chapter can serve as a stepping stone for quickly grasping the NLP technologies most relevant to RE. With both the NLP and RE landscapes constantly evolving, our hope is to maintain this chapter as a living document, continuously integrating emerging trends and pertinent techniques for automated requirements analysis.

\sectopic{Acknowledgements.}The first author is grateful for the financial support provided by the Natural Sciences and Engineering Research Council of Canada (NSERC) through the Discovery and Discovery Accelerator programs.

\setlength{\bibsep}{0pt plus 0.3ex} 

\bibliography{cited}

\end{document}